\input harvmac
\noblackbox
\overfullrule=0pt
\def\Title#1#2{\rightline{#1}\ifx\answ\bigans\nopagenumbers\pageno0\vskip1in
\else\pageno1\vskip.8in\fi \centerline{\titlefont #2}\vskip .5in}

%
%
\def\sq2{\sqrt{2}}

\def\s42{ 2^{-{1\over 4} } }

\def\cg{\cos (2\pi V)}
\def\sg{\sin (2\pi V)}
\def\cg2{\cos (\pi V)}
\def\sg2{\sin (\pi V)}
\def\cb2{\cos (\delta/2)}
\def\sb2{\sin (\delta/2)}

\font\cmss=cmss10 \font\cmsss=cmss10 at 7pt
\def\IZ{\relax\ifmmode\mathchoice
   {\hbox{\cmss Z\kern-.4em Z}}{\hbox{\cmss Z\kern-.4em Z}}
   {\lower.9pt\hbox{\cmsss Z\kern-.4em Z}}
   {\lower1.2pt\hbox{\cmsss Z\kern-.4em Z}}\else{\cmss Z\kern-.4emZ}\fi}

\def\s{{\rm sinh}^2\alpha }

\def\[{\left [}
\def\]{\right ]}
\def\({\left (}
\def\){\right )}
%
%
\lref\jp{J. Polchinski, hep-th/9510017.}
\lref\witb{E. Witten, hep-th/9510135.}
\lref\vgas{C. Vafa, hep-th/9511088.}
\lref\ghas{G. Horowitz and A. Strominger, hep-th/9602051.}
\lref\bsv{M. Bershadsky, V. Sadov and C. Vafa,
hep-th/9511222.}
\lref\vins{C. Vafa, hep-th/9512078.}
\lref\cvetd{M. Cvetic and D. Youm, hep-th/9507090.}
\lref\polc{J. Dai, R. Leigh and J. Polchinski, Mod. Phys.
Lett. {\bf A4} (1989) 2073.}
\lref\ascv{A. Strominger and C. Vafa, hep-th/9601029.}
\lref\hrva{P. Horava, Phys. Lett. {\bf B231} (1989) 251.}
\lref\bekb{J. Bekenstein, Phys. Rev {\bf D12} (1975) 3077.}
\lref\hawkb{S. Hawking, Phys. Rev {\bf D13} (1976) 191.}
\lref\wilc{P. Kraus and F. Wilczek, hep-th/9411219, Nucl. Phys.
{\bf B433} (1995) 403. }
\lref\send{A. Sen, hep-th/9510229, hep-th/9511026.}
\lref\cvet{M. Cvetic and A. Tseytlin, hep-th/9512031.}
\lref\kall{R. Kallosh, A. Linde, T. Ortin, A. Peet andA. van Proeyen, Phys. Rev. 
{\bf 
D46} (1992) 5278.}
\lref\lawi{F. Larsen and F. Wilczek, hep-th/9511064.}
\lref\bek{J. Bekenstein, Lett. Nuov. Cimento {\bf 4} (1972) 737,
Phys. Rev. {\bf D7} (1973) 2333, Phys. Rev. {\bf D9} (1974) 3292.}
\lref\hawk{S. Hawking, Nature {\bf 248} (1974) 30, Comm. Math. Phys.
{\bf 43} (1975) 199.}
\lref\cama{C. Callan and J. Maldacena, hep-th/9602043.}
\lref\sen{A. Sen, hep-th/9504147, Mod. Phys. Lett. {\bf A10} (1995) 2081.}
\lref\fks{S. Ferrara, R. Kallosh and A. Strominger, hep-th/9508072,
Phys. Rev. {\bf D 52}, (1995) 5412 .}
\lref\spn{J. Beckenridge, R. Myers, A. Peet and C. Vafa, hep-th/9602065.}
\lref\vbd{J. Beckenridge, D. Lowe, R. Myers, A. Peet, A. Strominger 
and C. Vafa, to appear.}
\lref\bk{R. Kallosh and B. Kol, hep-th/9602014.}
\lref\asop{A. Strominger, hep-th/9512059.}
\lref\twn{P. Townsend, hep-th/9512062.}
\lref\yz{S.-T. Yau and E. Zaslow, hep-th/9512121.}

%
\Title{\vbox{\baselineskip12pt
\hbox{hep-th/9603060}}}
{\vbox{
\centerline {Statistical Entropy of Four-Dimensional} 
\centerline{Extremal Black Holes}  }}
\centerline{Juan M. Maldacena$^{\natural}$ and Andrew Strominger$^\dagger$}
\vskip.1in
\centerline{$^\dagger$\it Department of Physics, University of California,
Santa Barbara, CA 93106, USA}
\vskip.1in
\centerline{$^{\natural}$\it Joseph Henry Laboratories, Princeton University,
Princeton, NJ 08544, USA}
\vskip1in
\centerline{\bf Abstract}

String theory is used to count microstates of four-dimensional extremal black 
holes in 
compactifications with $N=4$ and $N=8$ supersymmetry. The result agrees for 
large
charges with the Bekenstein-Hawking entropy. 
\Date{}
%

Recently it has been shown \refs{\ascv \cama \ghas \spn -\vbd } that string 
theory can, in 
some special cases,  provide a 
statistical  derivation of the Bekenstein-Hawking 
entropy\refs{\bek,\hawk},
by representing the black holes as bound states of D-branes and strings. The 
statistical 
entropy is the logarithm of the bound state degeneracy, which was counted using 
D-technology introduced in \refs{\polc,\hrva,\jp}.
Curiously the results so far have been limited to five dimensions. The reason 
for 
this 
is that four-dimensional black holes with nonzero horizon area can not be 
constructed 
from D-branes alone. Another type of object such as a symmetric fivebrane or 
Kaluza-
Klein monopole is required, and further technology is needed. In this paper we 
will  find 
the missing piece of technology in references \refs{\asop,\twn} and use it to 
compute the 
statistical entropy of certain four-dimensional extremal black holes in
 $N=4$ and $N=8$ supergravity  
theories. 
The result agrees with the Bekenstein-Hawking entropy, which was computed in a 
special 
$N=4$ case in \kall, more generally for $N=4$ in \refs{\cvetd, \cvet} 
and for $N=8$ in  
\bk.

The statistical entropy of four-dimensional black holes has been recently 
analyzed in 
\lawi\ with methods seemingly quite different from those used herein. 
It would be very 
interesting to understand the relation between the two approaches. 

The required modification of \ascv\ is rather simple and this presentation will 
be 
accordingly 
brief. Let us begin by rederiving the result of \ascv\ in a T-dualized picture 
with 
one extra $\hat S^1$-compactified dimension. Consider type IIA string 
theory on $X=Y\times S^1\times \hat S^1$, where $Y$ is 
$T^4$ for the $N=8$ case and $K3$ for the $N=4$ case. 
A  dual description of the D-brane configuration in \ascv\  
(obtained by T-dualizing along 
$\hat S^1$) 
consists of  $Q_6$ sixbranes wrapping $X$, $Q_2$ twobranes wrapping 
$S^1\times \hat S^1$, and right-moving momentum $n$ along the $ S^1$. 
We take $n, Q_2 >>1$. The twobranes 
are marginally bound to the sixbranes \refs{\send \vgas -\vins}. For 
$Q_6=1$ the momentum is carried by massless, right-moving modes of 
$(2,2)$ open strings
that end on the 
twobranes.  It is sufficient to 
consider the case $Q_6=1$ because duality implies the 
results can depend only on 
the 
product $Q_2Q_6$. (This has been explicitly verified in 
some cases \refs{\send \vgas \bsv \vins -\yz}.) BPS excitations 
of these $(2,2)$ strings correspond to 
transverse motion of the $Q_2$ twobranes within $Y$ (and 
the sixbrane).\foot{Since the two branes are separated in 
$Y$ the $(2,2)$ open strings going between different twobranes 
are massive and do not contribute to the extremal entropy 
as in \ascv . 
$(2,6)$ strings also do not contribute in this case ($Q_6=1$) 
because of charge confinement.
} Because $Y$ is four-dimensional this means 
there are $4Q_2Q_6$ bosons and their $4Q_2Q_6$ 
fermionic superpartners available 
to 
carry the momentum.\foot{We suppress here the anomalous shift of $Q_2$ for 
$K3$ \refs{\bsv,\vins} which is subleading for large $Q_2$.} The number of BPS-
saturated states 
of this system as a function of $Q_2, Q_6$ and $n$ follows from the standard 
$(1+1)$-dimensional entropy formula
\eqn\std{S=\sqrt{\pi (2N_B+N_F)EL\over 6},}
where $N_B~~(N_F)$ is the number of species of right-moving bosons (fermions), 
$E$ is the total energy and $L$ is the size of the box. Using $N_B=N_F=4Q_2Q_6$ 
and $E=2\pi n/L$, we find the $L$-independent result for the large $n$ 
thermodynamic 
limit \ascv
\eqn\setn{S_{stat}=2\pi\sqrt{Q_2Q_6n}.}

The Bekenstein-Hawking entropy was computed from the 
corresponding four dimensional 
 extremal black hole solutions in \refs{\kall,\cvetd,\cvet,\bk}.
The result, in our notation, for either $N=4$ or $N=8$, is 
\eqn\bhen{S_{BH}=2\pi\sqrt{Q_2Q_6nm}.}
The integer $m$ here is 
the axion charge carried by a symmetric fivebrane which wraps 
$Y\times S^1.$\foot{To facilitate comparison with \refs{\cvetd, \cvet}, we 
note that under type II-heterotic duality an $m$-wound symmetric fivebrane 
together with momentum $n$ becomes a fundamental heterotic string with 
$(winding, momentum)=(m,n)$ around $S^1$. The twobranes and sixbranes become the 
magnetic 
heterotic $S$-duals of a fundamental heterotic string with 
$(winding, momentum)=(Q_2,Q_6)$ associated to the $(20,4)$ part of the Narain 
lattice.} 
Since that charge is absent in this 
$\hat S^1$ 
compactification of the configuration of \ascv, $S_{BH}=0$. This is not a 
contradiction 
because  
in four dimensions $S_{BH}$ as computed from the leading low energy effective 
action 
always scales like $(charge)^2$, in contrast to five dimensions where it scales 
like 
$(charge)^{3/2}$. Since \setn\ scales like $(charge)^{3/2}$, it appears at 
leading
 order in 
five 
dimensions but is an invisible subleading correction in four\foot{
In fact the four dimensional solution with $m=0$ contains 
scalar fields that blow up  at the horizon, rendering the classical
geometry at the horizon singular.}. 

In order to get  a nonzero area in four dimensions, we must add $m$ fivebranes 
wrapping 
$Y\times S^1$. These $m$ fivebranes can be located anywhere on the $\hat S^1$. 
Each twobrane must intersect all $m$ fivebranes along the $S^1$.
 The effect of this was explained in 
\refs{\asop,\twn}. A twobrane can break and the ends 
separate (in $Y$) when it  crosses a fivebrane. 
Hence the $Q_2$ toroidal twobranes break 
up into $mQ_2$ 
cylindrical twobranes, each of which is bounded by a pair of fivebranes. The 
momentum-carrying open strings now carry an extra label describing which pair of 
fivebranes 
they lie in between. The number of species becomes $N_B=N_F=4mQ_2Q_6$. Inserting 
this into \std\ together with $E=2\pi n/L$ we obtain
\eqn\setn{S_{stat}=2\pi\sqrt{Q_2Q_6nm},}
In agreement with the semiclassical result \bhen\ for $S_{BH}$.

For the $N=4$ case 
 there are, in general,  28 electric charges $\vec Q$ and 28 magnetic 
charges $\vec P$ 
which 
lie in the $(22,6)$ Narain lattice. In our notation $2Q_2Q_6=\vec P^2$ and 
$ 2 nm=\vec Q^2$. Duality implies that the entropy depends only on
$\vec P^2, ~\vec Q^2$ and $\vec Q \cdot \vec P$. The general formula for the 
Bekenstein 
Hawking entropy is \refs{\cvetd,\cvet}
\eqn\cvf{
S_{BH}=\pi \sqrt{ \vec P^2 \vec Q^2-(\vec Q \cdot \vec P)^2}.
}
For our example the last term vanishes. It would be interesting to construct a
more general example for which this last term does not vanish, and so verify the
 general 
formula.

\vskip.2in

{\bf Acknowledgements}

We would like to thank  C. Callan, G. Horowitz, F. Larsen,  J. Polchinski 
and C. Vafa for useful 
discussions. After completing this work it has come to our attention 
that related ideas are being pursued by C. Johnson, R. Khuri and R. 
Myers. 
The research of A.S. is supported in part by DOE grant DOE-91ER40618
and the research of J.M. is supported in part by DOE grant DE-FG02-91ER40671.

\listrefs
\end